\documentclass[aps,preprint]{revtex4}
\usepackage{amssymb}
\usepackage{amsfonts}
\usepackage{amsmath}
\usepackage{graphicx}

\begin{document}

\preprint{}
\title[Heat transfer in metal films]{Low-temperature electron-phonon heat
transfer in metal films. }
\author{S. Cojocaru}
\affiliation{Horia Hulubei National Institute For Physics And Nuclear Engineering,
RO-077125, Magurele, Romania}
\author{D. V. Anghel}
\affiliation{Horia Hulubei National Institute For Physics And Nuclear Engineering,
RO-077125, Magurele, Romania}

\begin{abstract}
We consider the deformation potential mechanism of the electron-phonon
coupling in metal films and investigate the intensity of the associated heat
transfer between the electron and phonon subsystems. The focus is on the
temperature region below dimensional crossover $T<T^{\ast }$ where the
thermally relevant vibrations are described in terms of a
quasi-two-dimensional elastic medium, while electron excitations behave as a
three-dimensional Fermi gas. We derive an explicit expression for the power $%
P\left( T\right) $ of the electron-phonon heat transfer which explains the
behavior observed in some experiments including the case of metallic film
supported by an insulating membrane with different acoustic properties. It
is shown that at low temperatures the main contribution is due to the
coupling with Lamb's dilatational and flexural acoustic modes.
\end{abstract}

\maketitle

\section{Introduction}

\begin{figure}[tbph]
\begin{center}
\includegraphics[height=6.2cm,width=7.6cm]{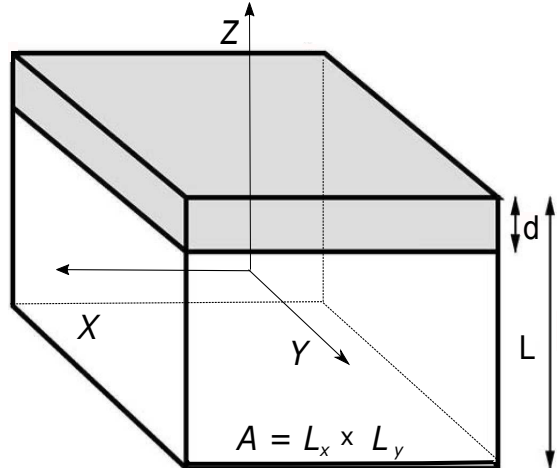}
\end{center}
\caption{Insulating membrane of thickness $L-d$ covered with a metal film
(gray) of thickness $d$ and surface area $A$.}
\label{fig1}
\end{figure}

In modern electronic devices the nanoscale miniaturization and sub-Kelvin
temperatures are quite common, however the physical phenomena taking place
in such conditions are far from complete understanding and attract a great
deal of research interest. In this work we address an aspect of the
electron-phonon interaction in confined systems related to heat transfer
between electrons and phonons. This is an open problem in the fundamental
sense with a direct connection to current research activity and important
applications in a variety of fields from nanoelectronics to astrophysics.
For instance, in a recent paper \cite{Nguyen} the principle of electronic
cooling \cite{Muhonen} has been used to realize the \textquotedblleft
coolest microfridge\textquotedblright\ reaching a record temperature of less
than $30$ mK. In a typical setup a cooled metal part is suspended or mounted
on an insulating support layer in contact with superconductors forming two
symmetrically biased NIS (normal metal-insulator-superconductor) tunnel
junctions. In this setup \textquotedblleft hot\textquotedblright\ electrons
from above the Fermi level are evacuated from the normal metal island, while
\textquotedblleft cold\textquotedblright\ electrons are injected below the
Fermi level. Such microdevices can be mounted directly on a chip for cooling
qubits or ultrasensitive low-temperature detectors, e.g., bolometers or
calorimeters, where the biased NIS tunnel junctions can also be used for
precision thermometry down to milli-Kelvin temperatures \cite{Feshchenko}.
An important physical phenomenon controlling the cooling power is the heat
transfer between electrons and phonons mediated by their coupling, $H_{e-p}$%
, when phonons are emitted and absorbed by electrons. When electrons are
heated by an external source in a stationary regime one can assume their
energy distribution to be characterized by a temperature $T_{e}$ while the
distribution of phonons corresponds to some lower temperature $T_{p}.$ In
many situations the temperature gradients are sufficiently small so that we
also assume that space variation of $T_{e}$ and $T_{p}$ can be neglected.
When both subsystems are bulklike (three-dimensional) the rate $P$ at which
electron energy is transferred to phonons has been obtained by considering
the deformation-potential mechanism of electron-phonon coupling, which
relates the local density fluctuation to the variation of the Fermi energy 
\cite{ziman}, and $P\left( T\right) $ has been shown to vary as $T^{\ 5}$ at
low temperatures (see, e.g., \cite{Kaganov, Allen, Wellstood}):%
\begin{equation}
P=\Sigma V_{el}\left( T_{e}^{5}-T_{p}^{5}\right) .  \label{power}
\end{equation}%
Here $V_{el}$ is the volume of the metal and $\Sigma $ depends on the
electron-phonon coupling and other properties of the sample. This form has
been derived for simple metals in the case when disorder is not strong ($%
ql>>1$, where $q$ is the phonon wave vector and $l$ is the electron mean
free path) and which is also assumed in the present work. It should be
mentioned that for disordered films a form with a stronger ($T^{6}$)
low-temperature behavior has been found \cite{Sergeev, Karvonen2}. The
dependence in Eq. (\ref{power}) was confirmed in many experimental
situations and is a standard formula assumed for the analysis of
experimental data, e.g., \cite{Nguyen, Feshchenko}. However, the finite
thickness of a film $L$ eliminates the possibility of longer waves to
propagate in this direction. Consequently, when temperatures fall below the
dimensional crossover threshold $T<T^{\ast }$ $\simeq c\hbar /\left(
k_{B}L\right) $, where $c$ is the sound velocity, the wavelength of the
thermally relevant phonons becomes longer than $L$ and we may treat the
phonon subsystem in terms of a confined elastic medium. Respectively,
thermal properties, including the electron-phonon heat transfer, are
dominated by the vibrational eigenmodes corresponding to such a
quasi-two-dimensional geometry. For values of $L$ of the order of 100 nm and
for sound velocities of the order of 10 km/s, the value of $T^{\ast }$ is of
the order of 1 K. Therefore size related effects in electron-phonon systems
have become an important part of the physics at the nanoscale, e.g., \cite%
{Stroscio, Cleland}. In a number of experimental studies it has been found
that the temperature dependence is best represented by the $T^{x}$ with
significantly lower values of $x$ \cite{Karvonen, Karvonen1, DiTusa}. On the
other hand, a theoretical investigation of the surface effects for a
half-space geometry \cite{Qu}, including the surface specific Rayleigh
phonon modes, has shown that the value of $x$ is actually larger than $5$
and at sufficiently low temperatures it exceeds $6$. Nevertheless, the
growth of the exponential $x$ with decreasing temperature has been later
qualitatively confirmed in some experiments when metallic films were
deposited on bulky substrates \cite{Underwood, Karvonen1}. For a
quasi-one-dimensional geometry (metallic nanowire) the model has been
studied theoretically in \cite{Hekking} where the $T^{3}$ analog of Eq. (\ref%
{power}) has been obtained. Although in \cite{Muhonen09} it was argued that
for Al nanowires with 65$\times $90-nm cross-section a better fit is
achieved with the standard exponential $x=5,$ the results are still
inconclusive since no dimensional crossover was observed around the
anticipated temperature, 0.45 K. In contrast, a clear indication of a
quasi-two-dimensional crossover in electron-phonon dominated heat flow with
a distinct power law ($x<4.5$ ) has been reported in Refs. \cite{Karvonen,
Karvonen1}. Remarkably, the respective samples had also a strongly enhanced
density of the heat flux compared to the thicker samples, which remained in
the \textquotedblleft bulk\textquotedblright\ regime and did not show a
crossover behavior for the considered temperature interval. Typically, the
metallic film is either deposited on an insulating membrane, Fig.\ref{fig1},
or suspended on top of superconducting electrodes \cite{Nguyen}.

The phonon spectrum of a quasi-two-dimensional system is quite different
from that of a bulk. Vibrational eigenstates of such a slablike structure
can not be separated into longitudinal and transverse waves; instead in the
elastic continuum approximation the spectrum is given by the Lamb eigenmodes
(in addition to shear waves) actually representing a mixture of both \cite%
{Auld}. The electron scattering by Lamb phonon modes has been studied
earlier for semiconductor quantum wells (QWs) in \cite{Bannov}, \cite%
{Stroscio} and in a double heterostructure QW including the piezoelectric
coupling in \cite{Glavin}. It has been shown that the scattering rate $\tau
^{-1}$ is dominated by coupling to the flexural acoustic mode (e.g., $\tau
^{-1}\sim T^{5/2},T^{7/2}$) due to its characteristic quadratic dispersion
and high density of states, in contrast to the \textquotedblleft
standard\textquotedblright\ linear dispersion of the, e.g., dilatational
Lamb mode, which has a negligible contribution ( e.g., $\tau ^{-1}\sim T^{6}$%
). In the present work the electron excitations in the metal film are
treated as a three-dimensional Fermi gas interacting with the
quasi-two-dimensional phonon subsystem. Thus, electrons are described by the
parabolic dispersion with an effective electron mass $\epsilon _{\mathbf{k}%
}=\hbar ^{2}\mathbf{k}^{2}/2m$ [notations for the components of the wave
vector $\mathbf{k=}\left( \mathbf{k}_{||},k_{z}\right) $ correspond to Fig.%
\ref{fig1}] and a plane-wave function $\Psi _{\mathbf{k}}\left( \mathbf{r,t}%
\right) =\exp \left( i\mathbf{kr}-i\epsilon t/\hbar \right) /\sqrt{V_{e}}%
=\psi \left( k_{z};z\right) \exp \left( i\mathbf{k}_{||}\mathbf{r}%
_{||}-i\epsilon t/\hbar \right) /\sqrt{A},$ where the electronic volume is $%
V_{e}=dA$ and $\psi \left( k_{z};z\right) $ is given in the next Section. We
will use alternatively either the cylindrical coordinate system with $k_{z}$
normal to the film and the in-plane vector $\mathbf{k}_{||}$ at angle $\phi $%
, or the spherical system with the two angles denoted as $\theta $ and $%
\varphi $ and $k=\left\vert \mathbf{k}\right\vert $. It should be mentioned
that in ultrathin metal films electron confinement can result in the
formation of the quantum well state with $\psi \left( k_{z};z\right) $ $\sim
\sin \left( k_{z}z\right) $ and $k_{z}=n\pi /d$, where the quantum numbers $%
n=1,2,3...$ correspond to the electron quasi-two-dimensional energy
sub-bands. However, in contrast to semiconductors, the QW state is more
difficult to observe in metal films thicker than a few nm (e.g., \cite{Mah}%
), because the electron de Broglie wavelength is comparable to the
interatomic distance and signatures of size quantization are easily smeared
out by film imperfections of the same length scale, e.g., surface roughness.
Thus, the plane-wave form of $\psi \left( k_{z};z\right) $ is an
approximation, which is meant to describe the situation of a not very thin
film where electron band structure can still be viewed as bulklike, while
the phonon spectrum is dominated by quasi-two-dimensional modes. An
important distinctive feature of this situation is that, although the
phonons propagate parallel to the plane of the film, they can nevertheless
produce electron scattering with a change of momentum in the direction
normal to the plane, i.e., $\hbar k_{z}^{\prime }\neq \hbar k_{z}.$ This
paradox is due to the displacement field pattern characteristic of the Lamb
waves (see below) which allows electrons to couple both to longitudinal
(in-plane) and transverse (out-of -plane) components of the vibrations (see
also the discussion in \cite{Lindenfeld} for the case of a nanowire). The
effect pertains primarily to the flexural modes and leads to a non-trivial
modification of the heat transfer.

In an often used experimental setup the metallic strip is in contact with an
insulating support membrane which can modify both the phonon spectrum and
the flow of the heat produced in the metal and transmitted through the
boundaries (e.g., \cite{Koppinen}). Thus, the boundary can give rise to
interface guided Stoneley phonon modes\cite{Brekh, Rose}. However, for a
solid-solid boundary the conditions on the parameters of the media
(densities and sound velocities) required for the existence of Stoneley
modes are very restrictive (see, e.g., Ref. \cite{Cheeke}). We return to
this issue in the next chapter. The heat transfer can also depend on the
coupling of phonons in the film to their own bath; e.g., in \cite{Pascal,
Underwood} it has been found that the distribution of phonons available for
interaction with electrons in metal films can remain relatively unaffected
by the substrate. In general, due to a mismatch of properties on the
interface between two materials the phonons will scatter and produce a
thermal boundary resistance; the associated Kapitza heat flow then depends
on the difference of phonon temperatures of the two materials and is usually
given by $\kappa \left( T_{p,1}^{4}-T_{p,2}^{4}\right) $ \cite{swartz, gia}.
There are other effects that can contribute to the heat transport, e.g.,
related to the operational principle of electron microcoolers, when there
exists a heat backflow from the superconductor to the metal island \cite%
{Heik}.

Below we will consider the heat flux derived from the electron-phonon
coupling for the structure shown in Fig.\ref{fig1} by first assuming
homogeneous elastic properties of the compound slab of volume $V_{p}=L\times
A$ \cite{Glavin}, its total mass $M,$ and mass density $\rho =M/V_{p}$. The
case of a suspended metallic film corresponds to the condition $L=d$. To
account for the modification of the phonon spectrum when the metal film is
deposited on the insulating membrane with acoustic characteristics different
from the metal, one can consider different models of their bonding (see,
e.g., \cite{Xiao}), however we assume that the contact between the two media
is rigid. It should be stressed that knowledge of the phonon spectrum is not
sufficient when considering the coupling to electrons and one should also
determine the properly normalized amplitudes of the phonon field.

\section{Electron-phonon coupling at low temperatures}

We define the rectangular coordinate system in such a way that $z=\pm L/2$
corresponds to the top and bottom surfaces of the slab. Elastic vibrations
are described by the vector field of relative displacements $\mathbf{u}=%
\mathbf{u}\left( \mathbf{r}\right) $ \cite{Auld} expanded in the series of
quantized eigenmodes of the continuum elasticity equation for the vibrations
of a rectangular plate \cite{Bannov}:%
\begin{equation}
\mathbf{u}\left( \mathbf{r}\right) =\sum_{\eta ,\mathbf{q}_{||}}\sqrt{\frac{%
\hbar }{2\rho A\omega _{\eta }}}\left[ a_{\eta }(\mathbf{q}_{||})+a_{\eta
}^{+}(-\mathbf{q}_{||})\right] \mathbf{w}_{\eta }(\mathbf{q}_{||},z)\exp
\left( i\mathbf{q}_{||}\mathbf{r}_{||}\right) .  \label{1}
\end{equation}%
Here $a_{\eta }^{+}(\mathbf{q}_{||})$ and $a_{\eta }(\mathbf{q}_{||})$ are
phonon creation and annihilation operators, and $\omega _{\eta }$ is the set
of normal vibration frequencies corresponding to the branches ($\xi $) of
the three types ($\alpha $) of eigenmodes $\eta =(\alpha =\left\{
h,d,f\right\} ,\xi =1,2,...)$, where $h$ is horizontal shear, $d$ is the
dilatational mode, and $f$ is the flexural mode. The quantum amplitudes $%
\mathbf{w}_{\eta }(\mathbf{q}_{||},z)$ are orthonormalized over the
thickness $L$%
\begin{equation}
\int_{-L/2}^{L/2}\mathbf{w}_{\eta }(\mathbf{q}_{||},z)^{\dag }\mathbf{w}%
_{\eta ^{\prime }}(\mathbf{q}_{||},z)dz=\delta _{\eta ,\eta ^{\prime }}.
\label{2}
\end{equation}%
The deformation potential coupling (see, e.g., \cite{ziman}) $H_{e-p}=\frac{2%
}{3}E_{F}\int_{V_{e}}d^{3}\mathbf{r}\,\,\Psi ^{\dagger }(\mathbf{r})\,\Psi (%
\mathbf{r})\,\nabla \cdot \mathbf{u}\left( \mathbf{r}\right) $ (where $%
E_{F}=\hbar ^{2}\mathbf{k}_{F}^{2}/2m$ is the Fermi energy; see also the
discussion for Cu in \cite{Qu}) is then determined by the divergence of the
displacement vector and takes the following form in the second quantization:%
\begin{equation}
H_{e-p}=\sum_{\mathbf{k_{||},\mathbf{q}_{||},}\eta ,k_{z},k_{z}^{\prime }}%
\left[ g_{\eta ,\mathbf{\mathbf{q}_{||}}}^{k_{z},k_{z}^{\prime }}\,c_{%
\mathbf{k_{||}+\mathbf{q}_{||},}k_{z}^{\prime }}^{\dagger }c_{\mathbf{k_{||},%
}k_{z}}\,a_{\eta }(\mathbf{q}_{||})+\left( g_{\eta ,\mathbf{\mathbf{q}_{||}}%
}^{k_{z},k_{z}^{\prime }}\right) ^{\ast }c_{\mathbf{k_{||}-\mathbf{q}_{||},}%
k_{z}^{\prime }}^{\dagger }\,\,c_{\mathbf{k_{||},}k_{z}}a_{\eta }^{\dagger }(%
\mathbf{q}_{||})\right] .  \label{3}
\end{equation}%
Here $c_{\mathbf{k_{||},}k_{z}}^{+}$ and $c_{\mathbf{k_{||},}k_{z}}$ are the
electron creation and annihilation operators; the electron-phonon matrix
elements are given by the expression: 
\begin{equation}
g_{\eta ,\mathbf{\mathbf{q}_{||}}}^{k_{z},k_{z}^{\prime }}=\frac{2}{3}E_{F}%
\sqrt{\frac{\hbar }{2\rho A\omega _{\eta }}}\int_{L/2-d}^{L/2}\psi ^{\ast
}\left( k_{z}^{\prime };z\right) \psi \left( k_{z};z\right) \left( i\mathbf{q%
}_{||}\cdot \mathbf{w}_{\eta }(\mathbf{q}_{||},z)+\frac{\partial w_{\eta
}^{z}(\mathbf{q}_{||},z)}{\partial z}\right) dz.  \label{3a}
\end{equation}%
Note that for the considered quasi-two-dimesnional geometry the momentum
conservation rule works only for the in-plane components of the wave
vectors, $\mathbf{k}_{||}^{\prime }=\mathbf{k}_{||}$ $\pm \mathbf{\mathbf{q}%
_{||}}$, and the electron-phonon coupling allows scattering with the change
of the $k_{z}$ component. The coordinate system is oriented in the plane as
shown in Fig.\ref{fig1} so that we choose $x$ to correspond to the
propagation direction of the wave $\mathbf{\mathbf{q}_{||}=}\left(
q,0,0\right) .$ Then from the displacement patterns of the three eigenmodes, 
$\mathbf{w}_{h}(q,z)=\left( 0,w_{h}^{y},0\right) $ and $\mathbf{w}%
_{d,f}(q,z)=\left( w_{d,f}^{x},0,w_{d,f}^{z}\right) ,$ one can easily see
that only the $d$ and $f$ modes couple to electrons. We further the mode
index whenever this does not cause confusion and then the amplitudes
resulting from the solutions of the elasticity equations for the free
surface boundary conditions can be represented as follows (see Refs. \cite%
{Dragos1, Dragos2}):%
\begin{equation*}
w_{d}^{x}=iq_{t}F_{d}\left[ 2q^{2}\cos \left( \frac{q_{t}L}{2}\right) \cos
\left( q_{l}z\right) +(q_{t}^{2}-q^{2})\cos \left( \frac{q_{l}L}{2}\right)
\cos \left( zq_{t}\right) \right] ,
\end{equation*}

\begin{equation}
w_{d}^{z}=qF_{d}\left[ -2q_{t}q_{l}\cos\left( \frac{q_{t}L}{2}\right)
\sin\left( q_{l}z\right) +(q_{t}^{2}-q^{2})\cos\left( \frac{q_{l}L}{2}%
\right) \sin\left( zq_{t}\right) \right] .  \label{4}
\end{equation}
and%
\begin{equation*}
w_{f}^{x}=iq_{t}F_{f}\left[ 2q^{2}\sin\left( \frac{q_{t}L}{2}\right)
\sin\left( q_{l}z\right) +(q_{t}^{2}-q^{2})\sin\left( \frac{q_{l}L}{2}%
\right) \sin\left( zq_{t}\right) \right] ,
\end{equation*}

\begin{equation}
w_{f}^{z}=qF_{f}\left[ 2q_{t}q_{l}\sin \left( \frac{q_{t}L}{2}\right) \cos
\left( q_{l}z\right) -(q_{t}^{2}-q^{2})\sin \left( \frac{q_{l}L}{2}\right)
\cos \left( zq_{t}\right) \right] .  \label{5}
\end{equation}%
The multipliers $F_{d}$ and $F_{f}$ are determined by the normalization
condition (\ref{2}). The above expressions are equivalent to Eqs. (10) -
(11) and (15) - (16) obtained in \cite{Bannov}, as can be easily checked by
taking into account that the auxiliary parameters ($q_{t},q_{l}$ ) satisfy
the eigen-frequency equations: 
\begin{equation}
\frac{-\ 4q^{2}q_{l}q_{t}}{(q^{2}-q_{t}^{2})^{2}}=\frac{\tan (q_{t}L/2)}{%
\tan (q_{l}L/2)},  \label{6}
\end{equation}%
for the dilatational mode, and 
\begin{equation}
\frac{-\ 4q^{2}q_{l}q_{t}}{(q^{2}-q_{t}^{2})^{2}}=\frac{\tan (q_{l}L/2)}{%
\tan (q_{t}L/2)},  \label{7}
\end{equation}%
for the flexural mode. The closure equation is secured by the
\textquotedblleft Snell law\textquotedblright :%
\begin{equation}
\omega =c_{l}\sqrt{q_{l}^{2}+q^{2}}=c_{t}\sqrt{q_{t}^{2}+q^{2}},  \label{8}
\end{equation}%
where $c_{l,t}$ are the longitudinal and transverse sound velocities of the
material with the Lame coefficients $\lambda $ and $\mu $ and mass density $%
\rho :$%
\begin{equation*}
c_{t}=\frac{\mu }{\rho };\ c_{l}=\frac{\lambda +2\mu }{\rho }:\ \ J\equiv
c_{t}^{2}/c_{l}^{2}<1/2.
\end{equation*}%
From the analysis of the solutions, Refs. \cite{Bannov} and \cite{Dragos2},
it follows that to obtain an explicit expression for the leading
low-temperature terms of $P$ it is sufficient to consider the lowest-energy
part of the phonon spectra (i.e., acoustical waves), so that the branch
index $\xi =1$ can now be dropped, $\omega _{\eta }\left( q\right) =\omega
_{\alpha =d,f}$. Then auxiliary parameters for the $f$ mode are both purely
imaginary ($q_{t.l}=ip_{t,l}$), while for the $d$ wave $q_{t}$ is real and $%
q_{l}=ip_{l}$ is imaginary. After lengthy but straightforward calculations
we obtain the explicit form of the solution of the above equations in the
long-wavelength approximation: 
\begin{equation}
\omega _{d}\simeq 2qc_{t}\sqrt{1-J},F_{d}^{-2}\simeq 16q^{6}\left(
3-4J\right) \left( 1-J\right) ^{2}L,\ \ q_{t}^{d}\simeq q\sqrt{3-4J},\ \ \
p_{l}^{d}\simeq q\left( 1-2J\right) .  \label{8a}
\end{equation}%
\begin{equation*}
\omega _{f}\simeq q^{2}Lc_{t}\sqrt{\left( 1-J\right) /3},\ F_{f}^{-2}\simeq
\left( q^{2}L\right) ^{6}L\left( 1-J\right) ^{2}/36,
\end{equation*}%
\begin{equation}
p_{t}^{f}\simeq -q+q^{3}L^{2}\left( 1-J\right) /6,\ \ \ p_{l}^{f}\simeq
-q+q^{3}L^{2}J\left( 1-J\right) /6.  \label{9}
\end{equation}%
These expressions can now be used for the calculation of the electron-phonon
matrix elements (\ref{3a}): 
\begin{equation}
\left\vert g_{\alpha ,\mathbf{\mathbf{q}_{||}}}^{k_{z},k_{z}^{\prime
}}\right\vert ^{2}=|F_{\alpha }|^{2}\left[ q_{t}q\left(
q_{l}^{2}+q^{2}\right) \right] ^{2}\frac{8\hbar E_{F}^{2}}{9\rho A\omega
_{\alpha }}%
\begin{Bmatrix}
\cos ^{2}\left( Lq_{t}/2\right) ,\alpha =d \\ 
\sin ^{2}\left( Lq_{t}/2\right) ,\alpha =f%
\end{Bmatrix}%
S\left( k_{z},k_{z}^{\prime },\alpha ,q_{l}\right) ,  \label{9a}
\end{equation}%
where the overlap integral $S\left( k_{z},k_{z}^{\prime },\alpha
,q_{l}=ip_{l}\right) $ is 
\begin{equation}
S\left( k_{z},k_{z}^{\prime },\alpha ,q_{l}=ip_{l}\right) =\left\vert
\int_{L/2-d}^{L/2}\psi ^{\ast }\left( k_{z}^{\prime };z\right) \psi \left(
k_{z};z\right) 
\begin{Bmatrix}
\cosh \left( zp_{l}\right) ,\alpha =d \\ 
\sinh \left( zp_{l}\right) ,\alpha =f%
\end{Bmatrix}%
dz\right\vert ^{2}.  \label{9b}
\end{equation}%
As we have already mentioned, the scattering processes described by the
Hamiltonian (\ref{3}) do not require conservation of the $z$ component of
the electron wave vector. However, in the long wave limit the overlap
integrals can be approximated by taking $\cosh \left( zp_{l}\right) \simeq 1$
and $\sinh \left( zp_{l}\right) \simeq zp_{l}$ in (\ref{9b}) and using the
orthonormality of the $\psi \left( k_{z};z\right) $. The respective
expressions for the $d$ and $f$ modes simplify to: 
\begin{equation}
S\left( k_{z},k_{z}^{\prime },d,q_{l}=ip_{l}\right) =\delta
_{k_{z},k_{z}^{\prime }},  \label{s}
\end{equation}%
and%
\begin{equation}
S\left( k_{z},k_{z}^{\prime },f,q_{l}=ip_{l}\right) =\left\vert
p_{l}\right\vert ^{2}\left\vert \int_{L/2-d}^{L/2}z\psi ^{\ast }\left(
k_{z}^{\prime };z\right) \psi \left( k_{z};z\right) dz\right\vert ^{2}.
\label{a}
\end{equation}%
Thus, in the long-wavelength approximation the interaction with dilatational
modes effectively preserves the electron momentum $k_{z}$, unlike the
interaction with flexural modes. As discussed in the Introduction, we assume
the plane-wave expression for the $\psi \left( k_{z};z\right) $ function: 
\begin{equation}
\psi \left( k_{z};z\right) =\sqrt{\frac{1}{d}}\exp \left( ik_{z}\left(
z+d-L/2\right) \right) .  \label{FG}
\end{equation}%
This implies, as pointed out in Ref. \cite{Lindenfeld} for the case of a
nanowire, that the electron-phonon coupling containing integrals like that
in Eq. (\ref{a}) diverges with the thickness of the film. However, below it
will be seen that this divergence is removed by the proper normalization of
the phonon eigenmodes [$F_{d}$ and $F_{f}$ in Eqs. (\ref{8a}) and (\ref{9})]
in the volume of the sample $V_{p}$ and the electron-phonon matrix element (%
\ref{9a}) remains finite.

\section{Heat flux carried by Lamb modes}

We can now calculate the power function $P,$ i.e., the energy transferred
from hot electrons to phonons in a unit of time: 
\begin{equation}
P=2\sum_{\mathbf{k_{||},\mathbf{q}_{||},}\alpha ,k_{z},k_{z}^{\prime }}\hbar
\omega _{\alpha }\left( \Gamma _{\alpha ,k_{z},k_{z}^{\prime }}^{\mathrm{em}%
}\left( \mathbf{k_{||}}\rightarrow \mathbf{k_{||}}-\mathbf{\mathbf{q}_{||}}%
\right) -\Gamma _{\alpha ,k_{z},k_{z}^{\prime }}^{\mathrm{ab}}\left( \mathbf{%
k_{||}}\rightarrow \mathbf{k_{||}}+\mathbf{\mathbf{q}_{||}}\right) \right) ,
\label{15}
\end{equation}%
The emission and absorption rates $\Gamma $ are given by the golden rule:%
\begin{equation*}
\Gamma ^{\mathrm{em}}(\mathbf{k_{||}}\rightarrow \mathbf{k_{||}}-\mathbf{%
\mathbf{q}_{||}})=\frac{2\pi }{\hbar }\,\,\left\vert g_{\alpha ,\mathbf{%
\mathbf{q}_{||}}}^{k_{z},k_{z}^{\prime }}\right\vert ^{2}\left[ n_{p}(\hbar
\omega _{\alpha })+1\right]
\end{equation*}%
\begin{equation}
\times f(\epsilon _{\mathbf{k_{||},}k_{z}})[1-f(\epsilon _{\mathbf{k_{||}}-%
\mathbf{\mathbf{q}_{||},}k_{z}^{\prime }})]\,\,\delta (\epsilon _{\mathbf{%
k_{||},}k_{z}}-\epsilon _{\mathbf{k_{||}}-\mathbf{\mathbf{q}_{||},}%
k_{z}^{\prime }}-\hbar \omega _{\alpha }),  \label{16}
\end{equation}%
where electron ($e$) and phonon ($p$) indices identify the respective
temperature in the Bose distribution function $n_{p,e}(\hbar \omega _{\alpha
})=\left\{ \exp \left( \beta _{p,e}\hbar \omega _{\alpha }\right) -1\right\}
^{-1}$ with $\beta _{p,e}=1/k_{B}T_{p,e}$; in the Fermi distribution
function the chemical potential is replaced by the Fermi energy for the
considered low-temperature regime $f(\epsilon _{\mathbf{k}})=\left\{ \exp %
\left[ \beta _{e}\left( \epsilon _{\mathbf{k}}-E_{F}\right) \right]
+1\right\} ^{-1}$. The phonon absorption part of (\ref{15}), $-\ \Gamma
_{\alpha }^{\mathrm{ab}}(\mathbf{k}\rightarrow \mathbf{k}+\mathbf{\mathbf{q}%
_{||}}),$ is obtained from the emission term $\Gamma _{\alpha }^{\mathrm{em}%
}(\mathbf{k}\rightarrow \mathbf{k}-\mathbf{\mathbf{q}_{||}})$ by the
space-time inversion ($\mathbf{\mathbf{q}_{||}\longrightarrow -\mathbf{q}%
_{||}}$ and $\omega _{\alpha }\longrightarrow -\omega _{\alpha }$ ) using
the identity for the Bose distribution $n(-y)+1=-n(y).$ The power function $%
P\left( T\right) $ can then be cast in the form of a difference between
terms separately dependent on $T_{e}$ and $T_{p}$, as in Eq. (\ref{power}),
with the help of the following identity for the Fermi and Bose distribution
functions: 
\begin{equation}
f(x)\left[ 1-f\left( x-y\right) \right] =n_{e}(y)\left[ f\left( x-y\right)
-f(x)\right] .  \label{identity}
\end{equation}%
Summation over momenta in Eq.(\ref{15}) is replaced by integration in a
standard way. In calculating the integrals one can then switch to electron
density of states and carry out energy integration by using the identity:%
\begin{equation}
\int_{0}^{\infty }\left[ f\left( x-y\right) -f(x)\right] dx=k_{B}T_{e}\ln
\left( \frac{\exp \left( \beta _{e}y\right) +\exp \left( -\beta
_{e}E_{F}\right) }{1+\exp \left( -\beta _{e}E_{F}\right) }\right) ,
\label{17a}
\end{equation}%
where the variable $y$ is defined by the phonon energy $y=\pm \hbar \omega
_{\alpha }$ in the emission and absorption processes [see Eqs. (\ref{15})].
Taking into account that $\hbar \omega _{\alpha },k_{B}T<<$ $E_{F}$, the
exponentially small terms in the round brackets on the right-hand-side of
Eq. (\ref{17a}) can be neglected and we obtain the relation 
\begin{equation}
\int_{0}^{\infty }\left[ f\left( x-y\right) -f(x)\right] dx\simeq y,
\label{17b}
\end{equation}%
also used in \cite{Glavin}. However, a more physically transparent way is to
note that at low temperatures the electron scattering takes place near the
Fermi surface, i.e., $\hbar \omega _{\alpha },k_{B}T<<\epsilon _{\mathbf{k}%
}\sim $ $E_{F}$. This allows us to approximate the right-hand-side of Eq. (%
\ref{identity}) with $n_{e}(y)y\delta \left( \epsilon _{\mathbf{k}%
}-E_{F}\right) $, by using the known expression for the main term of the
expansion $f\left( x-y\right) -f(x)\simeq -y\partial f/\partial x\simeq $ $%
y\delta \left( \epsilon _{\mathbf{k}}-E_{F}\right) $ [see, e.g., Eq. (5.42)
in \cite{Rickayzen}]. It can also be seen that this approximation reproduces
the result in Eq. (\ref{17b}); recall that $x=\epsilon _{k}$. Then the
above-mentioned form of Eq. (\ref{15}) is easily obtained as follows:%
\begin{equation}
P=P_{0}\left( T_{e}\right) -P_{0}\left( T_{p}\right) .  \label{17}
\end{equation}%
Here%
\begin{equation*}
P_{0}\left( T_{e}\right) =\sum_{\mathbf{k_{||}\mathbf{,}}k_{z},k_{z}^{\prime
}\mathbf{,\mathbf{q}_{||},}\alpha =d,f}\frac{32\pi \hbar ^{2}E_{F}^{2}}{%
9\rho A}\omega _{\alpha }|F_{\alpha }|^{2}\left[ q_{t}q(q_{l}^{2}+q^{2})%
\right] ^{2}n(\hbar \omega _{\alpha }/k_{B}T_{e})\delta \left( \epsilon _{%
\mathbf{k}}-E_{F}\right)
\end{equation*}%
\begin{equation}
\times 
\begin{Bmatrix}
\cos ^{2}\left( Lq_{t}/2\right) ,\alpha =d \\ 
\sin ^{2}\left( Lq_{t}/2\right) ,\alpha =f%
\end{Bmatrix}%
S\left( k_{z},k_{z}^{\prime },\alpha ,q_{l}\right) \,\delta (\epsilon _{%
\mathbf{k_{||},}k_{z}}-\epsilon _{\mathbf{k_{||}}-\mathbf{\mathbf{q}_{||},}%
k_{z}^{\prime }}-\hbar \omega _{\alpha }).  \label{18}
\end{equation}%
Equation (\ref{18}) shows that in this leading order of the low-temperature
expansion phonons are emitted and absorbed by electrons on the Fermi surface.

Let us now consider the energy conservation condition imposed by the last $%
\delta $ function in Eq. (\ref{18}) on the cosine of the angle between the
electron and phonon wave vectors written in spherical coordinates ($\sin
\theta \cos \varphi $): 
\begin{equation}
\delta \left( \frac{\hbar ^{2}}{m}\left( k_{F}q\sin \theta \cos \varphi
\right) +\frac{\hbar ^{2}}{2m}\left( k_{z}^{2}-\left( k_{z}^{^{\prime
}}\right) ^{2}-q^{2}\right) -\hbar \omega _{\alpha }\right) .  \label{19}
\end{equation}%
For the dilatational mode ($\alpha =d$) we have $k_{z}^{\prime }=k_{z}$ from
Eq. (\ref{s}) and then from Eq. (\ref{18}) we obtain%
\begin{align*}
P_{0}^{FG,d}\left( T_{e}\right) & =\frac{32\pi \hbar ^{2}E_{F}^{2}}{9\rho A}%
\frac{Ad}{\left( 2\pi \right) ^{3}}\int_{0}^{\infty }k^{2}dk\int_{0}^{\pi
}\sin \theta d\theta \int_{0}^{2\pi }d\phi \frac{A}{\left( 2\pi \right) ^{2}}%
\left( \int_{0}^{\infty }\int_{0}^{2\pi }qdqd\phi _{d}\right) \\
& \times \omega _{d}|F_{d}|^{2}\left[ q_{t}q(q_{l}^{2}+q^{2})\right]
^{2}n_{e}(\hbar \omega _{d})\delta \left( \frac{\hbar ^{2}k^{2}}{2m}%
-E_{F}\right) \cos ^{2}\left( Lq_{t}/2\right)
\end{align*}%
\begin{equation}
\times \frac{m}{\hbar ^{2}kq}\delta \left( \sin \theta \cos \phi -\frac{q}{2k%
}-\frac{m\omega _{d}(q)}{\hbar kq}\right) .  \label{19a}
\end{equation}%
One can see that due to conservation of the $z$ component of the electron
momentum the phonon is actually emitted in the direction orthogonal to $%
\mathbf{k,}$ i.e., $\left\vert \sin \theta \cos \varphi \right\vert <<1.$
Indeed, we can estimate $m\omega _{d}/\hbar k_{F}q\sim $ $c_{t}/v_{F}$ $<<1$
and, since $k_{F}\sim \pi /a_{0}$ ($a_{0}$ is the lattice spacing), also $%
q/2k_{F}<<1.$ So, the $\mathbf{k}$ vector is allowed to rotate without
restrictions in the plane orthogonal to $\mathbf{\mathbf{q}_{||}}$, while $%
\mathbf{\mathbf{q}_{||}}$ in its turn is free to rotate in the plane of the
film. Thus, integration over the electron $\left( \theta ,\phi \right) $ and
phonon $\left( \phi _{d}\right) $ angles in Eq. (\ref{19a}) results in the
multiplier $\left( 2\pi \right) ^{2}$ (formal derivation is somewhat
lengthier and leads to the same conclusion). Integration over the length of $%
\mathbf{k}$ is trivial and can also be expressed in terms of the electron
density of states for the parabolic dispersion, $N\left( E_{F}\right)
=Admk_{F}/\left( \hbar \pi \right) ^{2}$. We finally obtain the following
expression for the $d$ mode:%
\begin{equation}
P_{0,d}\left( T_{e}\right) =\frac{32\pi m^{2}E_{F}^{2}}{9\rho \hbar ^{2}}%
\frac{Ad}{\left( 2\pi \right) ^{3}}\int_{0}^{\infty }\omega _{d}|F_{d}|^{2}%
\left[ q_{t}q(q_{l}^{2}+q^{2})\right] ^{2}n_{e}(\hbar \omega _{d})\cos
^{2}\left( Lq_{t}/2\right) dq,  \label{20}
\end{equation}%
which after substitution of the expressions in Eq. (\ref{8a}) into Eq. (\ref%
{20}) results in%
\begin{equation}
P_{0,d}\left( T_{e}\right) =\frac{\zeta \left( 4\right) }{12\pi ^{2}}\frac{%
V_{e}\left( k_{B}T_{e}\right) ^{4}k_{F}^{4}J^{2}}{\rho L\hbar
^{2}c_{t}^{3}\left( 1-J\right) ^{3/2}}.  \label{FG_S}
\end{equation}

Calculation of the heat current due to the flexural phonon modes is more
involved since $k_{z}$ is not conserved even in the long-wavelength
approximation and can significantly differ from $k_{z}^{^{\prime }}$, Eq. (%
\ref{a}). It is then convenient to express the power function (\ref{18}) in
the cylindrical coordinates:

\begin{align}
P_{0,f}\left( T_{e}\right) & =\frac{16E_{F}^{2}}{9\rho }\frac{Ad^{4}}{\left(
2\pi \right) ^{4}}\left( \frac{m}{\hbar }\right) ^{2}  \label{21} \\
& \times \int_{0}^{\infty }dqI\left( q\right) \left\vert p_{l}\right\vert
^{2}\omega _{\alpha }|F_{\alpha }|^{2}\left[ q_{t}q(q_{l}^{2}+q^{2})\right]
^{2}n(\hbar \omega _{q}/k_{B}T_{e})\sin ^{2}\left( Lq_{t}/2\right) .  \notag
\end{align}%
Here%
\begin{align}
I\left( q\right) & =\int_{0}^{k_{F}}\frac{dk_{||}}{\sqrt{k_{F}^{2}-k_{||}^{2}%
}}\int_{-k_{F}}^{k_{F}}dk_{z}\int_{-\infty }^{\infty }dk_{z}^{^{\prime }}
\label{I} \\
& \times \left[ \delta \left( k_{z}-\sqrt{k_{F}^{2}-k_{||}^{2}}\right)
+\delta \left( k_{z}+\sqrt{k_{F}^{2}-k_{||}^{2}}\right) \right] \left\vert
\int_{\sigma /2-1}^{\sigma /2}\exp \left( id\left( k_{z}-k_{z}^{\prime
}\right) z\right) zdz\right\vert ^{2}  \notag \\
& \times \int_{0}^{2\pi }d\phi \delta \left( \cos \phi +\frac{\left(
k_{z}^{2}-\left( k_{z}^{^{\prime }}\right) ^{2}-q^{2}\right) /2}{k_{||}q}-%
\frac{m\omega _{f}}{\hbar k_{||}q}\right) .  \notag
\end{align}%
In (\ref{I}) we have introduced the ratio $\sigma =L/d$ with the limit value 
$\sigma =1$ corresponding to the absence of the insulating membrane, i.e.,
to a purely metallic sample. The multiple integral $I\left( q\right) $ is
calculated analytically in the Appendix and its substitution into Eq. (\ref%
{21}), together with the solution (\ref{9}) for the flexural mode, leads to
the following expression:%
\begin{equation}
P_{0,f}\left( T_{e}\right) =\frac{V_{e}\left( k_{B}T_{e}\right)
^{4}d^{3}m^{2}J^{2}E_{F}^{2}}{4\sqrt{3}\pi ^{3}\rho L^{4}c_{t}^{3}\hbar
^{6}\left( 1-J\right) ^{3/2}}\left[ \left( \sigma -1\right)
^{2}\int_{0}^{\infty }x^{3}\left( -\ln x\right) n\left( x\right) dx\right.
\label{22}
\end{equation}%
\begin{equation*}
\left. +\left( \frac{1}{2}+\left( \sigma -1\right) ^{2}\ln \left( 25.53\frac{%
\hbar Lc_{t}\sqrt{\left( 1-J\right) /3}}{d^{2}k_{B}T_{e}}\right) \right)
\int_{0}^{\infty }x^{3}n\left( x\right) dx\right] .
\end{equation*}%
From Eq. (\ref{22}) we obtain the final result for the contribution of the
flexural modes: 
\begin{equation}
P_{0,f}\left( T_{e}\right) =\frac{0.0075V_{e}\left( k_{B}T_{e}\right)
^{4}k_{F}^{4}J^{2}}{\rho L\sigma ^{3}\hbar ^{2}c_{t}^{3}\left( 1-J\right)
^{3/2}}\left( \left( \sigma -1\right) ^{2}\ln \left[ 4.4\frac{\sigma
^{2}\hbar c_{t}\sqrt{1-J}}{Lk_{B}T_{e}}\right] +\frac{1}{2}\right) .
\label{23}
\end{equation}%
The argument in the square brackets is proportional to $T^{\ast }/T_{e}\sim
\hbar c/\left( Lk_{B}T_{e}\right) $, so that the log term is positive for
temperature in the interval below the crossover, i.e., where the description
of the phonon subsystem in terms of quasi-two-dimensional phonon confinement
is applicable. Note also that $P_{0,f}\left( T_{e}\right) $ does not vanish
even for a purely metallic slab ($V_{e}=V_{p}$ and $\sigma =1$) when
electrons experience an antisymmetric field created by the flexural
vibration mode [Eq. (\ref{a})] and the respective overlap function vanishes $%
S\left( k_{z}=k_{z}^{\prime },f,q_{l}=ip_{l}\right) =0$. This result
demonstrates the point made in the Introduction that the non-zero
contribution of the flexural modes to the heat transfer is due to scattering
processes with $k_{z}^{\prime }\neq k_{z}$. One can also see that the above
result scales with the surface of the sample ($P\thicksim A$) for a free
metallic film ($\sigma =1$), but for a composite structure ($\sigma >1$) the
geometry dependence becomes more complicated even for an acoustically
uniform medium.

To simplify the discussion we assume that the phonon temperature is much
lower than $T_{e}$, so that the total density of the heat transfer power $Q$
is obtained as the sum of just the two contributions in Eqs. (\ref{FG_S})
and (\ref{23}): 
\begin{equation}
Q=\left( P_{0,d}\left( T_{e}\right) +P_{0,f}\left( T_{e}\right) \right)
/V_{e}.  \label{23a}
\end{equation}%
Thus, for the considered case of an acoustically uniform metal-insulator
composite slab we obtain the following expression: 
\begin{equation}
Q=\frac{0.0075\left( k_{B}T_{e}\right) ^{4}k_{F}^{4}J^{2}}{\rho L\sigma
^{3}\hbar ^{2}c_{t}^{3}\left( 1-J\right) ^{3/2}}\left( 1.2\sigma ^{3}+\frac{1%
}{2}+\left( \sigma -1\right) ^{2}\ln \left( 4.4\frac{\hbar \sigma ^{2}c_{t}%
\sqrt{1-J}}{Lk_{B}T_{e}}\right) \right) .  \label{24}
\end{equation}%
Its generalization to the case of two acoustically inequivalent rigidly
bonded layers is much lengthier and will be presented in detail elsewhere. A
typical example corresponds to a Cu film deposited on a silicon-nitride
insulating membrane. Following the guidelines of the standard description of
layered elastic media (see, e.g., \cite{Brekh, Rose}), one can obtain the
analytic solutions for the acoustic branches of the vibrational modes and
respective normalization factors in the long-wavelength approximation. The
result is that solutions of the Stoneley type (i.e., interface guided waves
with the amplitude decreasing away from the interface) do not appear in the
long-wavelength limit and it is well justified to keep the two types of
modes considered above solely responsible for the low-temperature behavior
also in this case.

\begin{figure}[tbph]
\begin{center}
\includegraphics[height=6.2cm,width=7.6cm]{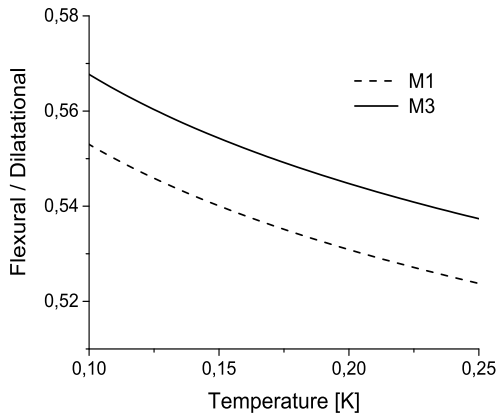}
\end{center}
\caption{The relative contribution $Q_{f}/Q_{d}$ , Eqs. (\protect\ref{28a})
and (\protect\ref{28b}), to the heat flux of the flexural versus
dilatational acoustic modes for the two samples $M1$ and $M3$ in\protect\cite%
{Karvonen, Karvonen1} (see text).}
\label{fig2}
\end{figure}

We identify the material parameters corresponding to Cu and silicon nitride
by the respective indices, $i=1$ and $2.$ The mass of the composite slab is $%
M$ and the ratio $M/A$ replaces the product $\rho L$ in Eq. (\ref{24}).
Then, with the additional notations%
\begin{equation}
R_{i}=\rho _{i}c_{t,i}^{2}\left( 1-J_{i}\right) ,\ \ \chi =\frac{1}{2}\left( 
\frac{R_{2}\left( L-d\right) ^{2}-R_{1}d^{2}}{R_{2}\left( L-d\right) +R_{1}d}%
\right) ,  \label{25}
\end{equation}%
\begin{equation*}
G=\frac{2\hbar }{d^{2}}\sqrt{\frac{\left( R_{1}d^{3}+R_{2}\left( L-d\right)
^{3}\right) /3-\chi ^{2}\left( R_{1}d+R_{2}\left( L-d\right) \right) }{\rho
_{1}d+\rho _{2}\left( L-d\right) }},
\end{equation*}%
the long-wavelength dispersion of the dilatational acoustic mode can be
written as%
\begin{equation}
\omega _{d}=2q\sqrt{\frac{R_{1}d+R_{2}\left( L-d\right) }{\rho _{1}d+\rho
_{2}\left( L-d\right) }}.  \label{26}
\end{equation}%
Equation (\ref{26}) reproduces the known result \cite{Xiao}. For the
flexural mode we find:%
\begin{equation}
\omega _{f}=2q^{2}\sqrt{\frac{\left( R_{1}d^{3}+R_{2}\left( L-d\right)
^{3}\right) /3-\chi ^{2}\left( R_{2}\left( L-d\right) +R_{1}d\right) }{\rho
_{1}d+\rho _{2}\left( L-d\right) }}.  \label{27}
\end{equation}%
It is easy to check that these expressions correctly reproduce the limit of
the acoustically uniform medium and amount to an effective renormalization
of the parameters in Eq. (\ref{24}) without qualitatively changing the
temperature dependence. Indeed, by carrying out the calculations within the
lines described in the previous case we obtain the following generalization
of Eqs. (\ref{22})$-$(\ref{23a}) for the power density function $Q\left(
T_{e}\right) =Q_{d}\left( T_{e}\right) +Q_{f}\left( T_{e}\right) $, where%
\begin{equation}
Q_{d}=\left( k_{B}T_{e}\right) ^{4}\frac{\zeta \left( 4\right) }{12\pi ^{2}}%
\frac{Ak_{F}^{4}J_{1}^{2}}{\hbar ^{2}M}\left( \frac{\rho _{1}d+\rho
_{2}\left( L-d\right) }{R_{1}d+R_{2}\left( L-d\right) }\right) ^{3/2},
\label{28a}
\end{equation}%
\begin{align}
Q_{f}& =\frac{\left( \kappa _{B}T_{e}\right) ^{4}}{\pi ^{3}3^{2}2^{7}}\frac{%
Ak_{F}^{4}J_{1}^{2}}{\hbar ^{2}M}\left( \left( 2\chi /d+1\right) ^{2}\left(
\ln \left( G/k_{B}T_{e}\right) +13.3\right) +3.247\right)  \label{28b} \\
& \times \left( \frac{d^{2}\left( \rho _{1}d+\rho _{2}\left( L-d\right)
\right) }{\left( R_{1}d^{3}+R_{2}\left( L-d\right) ^{3}\right) /3-\chi
^{2}\left( R_{2}\left( L-d\right) +R_{1}d\right) }\right) ^{3/2}.  \notag
\end{align}%
Figure \ref{fig2} shows the relative contribution of the two modes as given
by Eqs. (\ref{28a}) and (\ref{28b}) to the electron-phonon heat transfer ($%
Q_{f}/Q_{d}$) for the two samples $M1$ and $M3$ which have been interpreted
in Refs. \cite{Karvonen, Karvonen1} as demonstrating a crossover in the
sub-Kelvin region. The material parameters can be found in \cite{Karvonen,
Karvonen1, Qu}: $\rho _{1}=8940kg/m^{3},\rho
_{2}=3290kg/m^{3},c_{t,1}=2575m/s,c_{t,2}=6200m/s,J_{1}=0.27,J_{2}=0.36,k_{F}=1.\,\allowbreak 65\times 10^{10}m^{-1},L=30nm,A=600\times 300\left( \mu m\right) ^{2},d\left( M1\right) =14nm,d\left( M3\right) =19nm, 
$ to a good approximation $M/A\simeq \rho _{2}\left( L-d\right) $ by taking
into account that the surface of the Cu film in these experiments was
smaller than the supporting membrane.

We can see that the contribution of the flexural mode to the heat transfer
is comparable to the dilatational one and gains more \textquotedblleft
weight\textquotedblright\ towards lower temperatures due to the presence of
the log term in Eq. (\ref{28b}). Moreover, Fig.\ref{fig2} indicates that the
sample with a higher thickness ratio between the metal film and the
insulating membrane (i.e., a smaller geometric factor: $\sigma \left(
M3\right) =49/19<\sigma \left( M1\right) =44/14$) has also a higher value of
the power ratio $Q_{f}/Q_{d}$. Note also that the ratio $Q_{f}/Q_{d}$
reduces to a constant ($\simeq 0.41$) for the case of a suspended metallic
film, as can be easily seen from the Eqs. (\ref{FG_S}) and (\ref{23}) with $%
\sigma =1$.

\begin{figure}[tbph]
\begin{center}
\includegraphics[height=6.2cm,width=7.6cm]{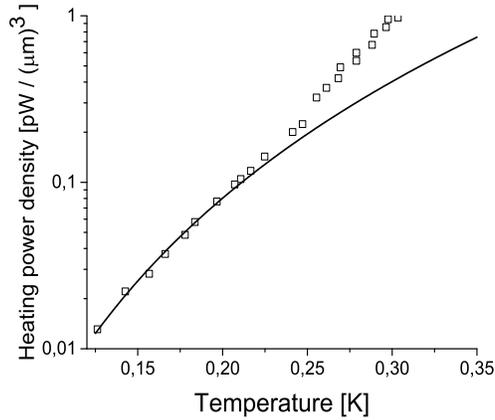}
\end{center}
\caption{The power density of the electron-phonon heat transfer for the $M1$
sample in \protect\cite{Karvonen} (squares) and the joint contribution due
to flexural and dilatational acoustical Lamb modes $Q=Q_{f}+Q_{d}$ as given
by Eqs. (\protect\ref{28a}) and (\protect\ref{28b}) (see text).}
\label{fig3}
\end{figure}

In Fig. \ref{fig3} our result is compared to the temperature dependence of
the total power density $Q=Q_{f}+Q_{d}$ for the sample $M1$ (Fig. 3 in \cite%
{Karvonen}) with the material parameters as given above. The effective value
of $k_{F}$ is obtained by fitting the low-temperature region of $Q(T)$ and
is slightly larger than $1.4\times 10^{10}$ m$^{-1}$ following from the
known value of the Fermi energy for copper, $7$ eV, if the effective mass $m$
is estimated from the electron heat capacity (e.g., \cite{Martin}) for the
simple isotropic parabolic dispersion. We mention that in \cite{Qu} the
deviation of the Fermi surface in noble metals from a simple spherical shape
has been studied in terms of surface averaged effective electronic
parameters and for Cu the estimated increase of the respective prefactor in
the electron-phonon power function is comparable to our result. 
At temperatures above $250$ mK the analytical curve starts to deviate from
the experiment as the crossover temperature is approached and higher-energy
branches of the Lamb modes spectrum should be taken into account. From the
Eqs.(\ref{28a}) and (\ref{28b}) it also follows that the $M3$ sample (not
shown in Fig.(\ref{fig3})) has a somewhat higher value of the total power
density $Q$ than $M1$ for the considered temperature region. The
quantitative comparison shows a good agreement with the results presented in
Fig. 4 of \cite{Karvonen} and in Fig. 3 of \cite{Karvonen1}.

\section{Conclusions}

We have obtained explicit expressions for the power density of the
electron-phonon heat transfer in a metallic film including the case when the
film is deposited on an insulating membrane with generally different
acoustic characteristics, Eqs. (\ref{28a}) and (\ref{28b}). The temperature
regime covered by the present analysis corresponds to low temperatures when
the phonon spectrum is dominated by quasi-two-dimensional modes of
vibration. The long-wavelength approximation which considers only the lowest
phonon branches is well justified at lower temperatures and has allowed us
to carry out the calculations analytically. Thus, for the specific example
considered above, e.g., Fig. \ref{fig3}, this description corresponds to
temperatures below\ $0.25$ K while the quasi-two-dimensional regime sets in
around\ 0.4-0.5 K \cite{Karvonen}. It turns out that the contributions of
the flexural and dilatational phonon modes to the heat flux are of the same
order of magnitude, as illustrated by Fig. \ref{fig2}.

For a suspended metallic film, $\sigma =1$, the heat current follows a $%
T^{4} $ dependence, i.e., $P=\Sigma _{2D}A\left( T_{e}^{4}-T_{p}^{4}\right) $%
, which appears to \textquotedblleft fit the pattern\textquotedblright\ of
the dimensionality dependence including the integer power $x$ in the $T^{x}$
temperature variation, namely, when the phonon subsystem corresponds to
either a three-dimensional bulk material with the power index $x=5$ as in
Eq. (\ref{power}), or to a two-dimensional or quasi-two-dimensional material
such as a single or bilayer graphene \cite{Viljas, kub} with $x=4$, or to a
quasi-one-dimesnional nanowire \cite{Hekking} with $x=3$. However, this
result could seem surprising in the context of previous works on some other
quasi-two-dimensional systems, e.g., \cite{Bannov, Glavin}, for the case of
a semiconductor quantum well, which would rather suggest a fractional value
of $x$ due to the peculiar quadratic dispersion of the flexural mode.
Fractional power, $T^{2.5}\ln T$, was also reported for the flexural modes'
contribution to electrical resistivity in free-standing graphene \cite{eros}%
. The value $x=4$ for the heat flux in graphene is actually due to electron
coupling with the dilatational phonons, since the \textquotedblleft
troublesome\textquotedblright\ flexural ones couple to electrons only in
second order in the displacement and can be disregarded for the graphene on
a substrate as well \cite{Viljas}. Moreover, essentially different
contributions of the phonon modes could be expected to result not only from
the linearity and non-linearity of the respective dispersion laws, but also
from the explicit presence of the size (thickness $L$) dependence in the
dispersion of the flexural mode [e.g., compare $\omega _{d}$ and $\omega
_{f} $ in Eqs. (\ref{8a}) and (\ref{9})]. Indeed, in the studies on phonon
transport and phonon heat capacity of a free-standing dielectric membrane 
\cite{Kuhn2, Fefelov} a striking difference between the thickness dependent
and non-dependent behavior has been derived for the two types of vibration.
In contrast, for the electron-phonon heat transfer we find a comparable
contribution for the two modes even when the metallic film is deposited on
an insulating membrane ($\sigma >1$), when the effect of sample geometry on
the power $P$ is more complicated than a simple scaling with surface area $A$
[see, e.g., Eqs. (\ref{FG_S}) and (\ref{23}) for the acoustically uniform
sample]. As we have shown such unexpected relative similarity in the
temperature and size (e.g., thickness) dependence results from several
physical factors, so that their combination differs from the cases studied
earlier. For a coupled electron-phonon system both the dispersions and the
amplitudes of the excitations play an important role. Respectively, a proper
normalization of the amplitudes is crucial. One can then see, e.g., from
Eqs. (\ref{8a}) and (\ref{9}), that the normalization of the phonon modes in
Eqs. (\ref{4}) and (\ref{5}), containing the multipliers $F_{d,f}$ and the
auxiliary parameters $q_{t,l}$, moderates to some extent the sharp
differences between the $d$ and $f$ dispersions. It is also clear that this
\textquotedblleft compensation\textquotedblright\ depends on specific
quantity considered, which in our case is the heat flux. Unlike the case of
graphene or that of a semiconductor quantum well, the electron excitations
are described in terms of three-dimensional Fermi gas with parabolic
dispersion. Respectively, the overlap of the electronic amplitudes in the
initial and scattered states [Eq. (\ref{9b})] is also a physical factor
which differs from the models considered before. As discussed in the text,
this overlap strongly discriminates between the $d$ and $f$ phonons and
contributes to the above-mentioned \textquotedblleft
compensation\textquotedblright\ as well.

For the case of a metallic film deposited on an insulating membrane the
flexural mode contribution to the power density of the electron-phonon heat
transfer acquires an additional logarithmic term of the form $T^{4}\ln T$
[Eq. (\ref{28b})], while that of dilatational modes keeps the $T^{4}$
dependence [Eq. (\ref{28a})]. As we have shown, such functional dependence
reproduces well the observed behavior, which was modeled in \cite{Karvonen}
with a power law $T^{x}$ with $x<4.5$. The dependence on the material
parameters for a metallic film with dielectric backing becomes more
complicated, especially for the acoustically non-uniform case. This point
can be illustrated by a more detailed analysis of the experimental work
cited above. Thus, one generally expects that reducing the dimensionality
would enhance the electron-phonon heat exchange. This is confirmed by Figs.
2 and 4 in \cite{Karvonen} for sufficiently thick samples to be considered
bulklike (e.g., $M5$ and $B1$) and which show a much lower power density $Q$
($=P/V_{e}$) than the thinner samples (like $M1$ and $M3$ discussed earlier,
with $M3$ being thicker than $M1$). Moreover, if the simple surface area
scaling of the power function ($P\sim A$) would be valid for these thinner
samples, then one would expect that $Q\left( M1\right) >Q\left( M3\right) $.
However, the power density for the thicker of the two ($M3$) is larger than
for the thinner one ($M1$) in the considered temperature region. This
non-trivial dependence on the material parameters is well reproduced by the
analytical expressions. Thus, the present analysis demonstrates that in thin
metal films both types of Lamb's phonon modes should be taken into account
on equal footing for a better understanding of the electron-phonon heat
transfer at low temperatures.

\begin{acknowledgments}
This work has been financially supported by UEFISCDI (Romania), Project No.
Idei-PCE 114/2011, and by ANCS, Project No. PN09370102.
\end{acknowledgments}

%\newpage
\setcounter{equation}{0} \renewcommand{\theequation}{A\arabic{equation}}

\section{APPENDIX}

Integration over the angle $\phi$ and over $k_{z}$ in (\ref{I}) reduces to a
tripple integral%
\begin{align}
& I\left( q\right) =\int_{0}^{k_{F}}\frac{dk_{||}}{\sqrt{k_{F}^{2}-k_{||}^{2}%
}}\int_{-\infty}^{\infty}dk_{z}^{^{\prime}}\frac{\theta\left( 1-\left( \left[
k_{F}^{2}-k_{||}^{2}-\left( k_{z}^{^{\prime}}\right) ^{2}\right]
/2k_{||}q-Dq/k_{||}\right) ^{2}\right) }{\sqrt{1-\left( \left[
k_{F}^{2}-k_{||}^{2}-\left( k_{z}^{^{\prime}}\right) ^{2}\right]
/2k_{||}q-Dq/k_{||}\right) ^{2}}}  \label{a1} \\
& \left( \left\vert \int_{\sigma/2-1}^{\sigma/2}\exp\left( id\left( \sqrt{%
k_{F}^{2}-k_{||}^{2}}-k_{z}^{\prime}\right) z\right) zdz\right\vert
^{2}+\left\vert \int_{\sigma/2-1}^{\sigma/2}\exp\left( id\left( \sqrt {%
k_{F}^{2}-k_{||}^{2}}+k_{z}^{\prime}\right) z\right) zdz\right\vert
^{2}\right) ,  \notag
\end{align}
where $\theta\left( x\right) $ is the Heaviside function and we have
introduced the constant%
\begin{equation*}
\ D=1/2+mLc_{t}\sqrt{\left( 1-J\right) /3}/\hbar>0\ ,
\end{equation*}
taking into account the dispersion of the flexural mode in (\ref{9}).

We further use the dimensionless variables%
\begin{equation}
x=\sqrt{k_{F}^{2}-k_{||}^{2}}/k_{F};\ \ y=k_{z}^{\prime}/k_{F};\ \ \
Q=q/k_{F}.  \label{a2}
\end{equation}
Eq. (\ref{a1}) then transforms into 
\begin{equation}
I\left( q\right) =2q\int_{0}^{1}dx\int_{-\infty}^{\infty}dy\frac {%
\theta\left( 4Q^{2}\left( 1-x^{2}\right) -\left( x^{2}-y^{2}-2DQ^{2}\right)
^{2}\right) }{\sqrt{4Q^{2}\left( 1-x^{2}\right) -\left(
x^{2}-y^{2}-2DQ^{2}\right) ^{2}}}\left( G^{-}+G^{+}\right) ,  \label{a3}
\end{equation}
where we have defined the functions%
\begin{equation}
G^{\mp}\equiv G\left( y\mp x\right) =\left\vert \int_{\sigma/2-1}^{\sigma
/2}\exp\left( idk_{F}\left( y\mp x\right) z\right) zdz\right\vert ^{2}.
\label{a4}
\end{equation}
We next consider the contribution $I^{-}\left( q\right) $ to $I\left(
q\right) $ corresponding to $G^{-}$ and switch to new variables%
\begin{equation}
u=y-x;\ \ v=x.  \label{a5}
\end{equation}
Then $I^{-}\left( q\right) $ takes the following form%
\begin{equation}
I^{-}=2q\int_{-\infty}^{\infty}du\left\vert
\int_{\sigma/2-1}^{\sigma/2}\exp\left( idk_{F}uz\right) zdz\right\vert
^{2}\int_{0}^{1}\frac {\theta\left( c-bv-av^{2}\right) dv}{\sqrt{c-bv-av^{2}}%
},  \label{a6}
\end{equation}
where $a,b$ and $c$ are functions of $u:$%
\begin{equation*}
a=4\left( Q^{2}+u^{2}\right) >0,
\end{equation*}%
\begin{equation*}
b=4u\left( u^{2}+2DQ^{2}\right) ,
\end{equation*}%
\begin{equation}
c=4Q^{2}-\left( u^{2}+2DQ^{2}\right) ^{2}.  \label{a7}
\end{equation}
Integration over $v$ gives%
\begin{equation}
\int_{0}^{1}\frac{\theta\left( c-bv-av^{2}\right) dv}{\sqrt{c-bv-av^{2}}}=%
\frac{1}{\sqrt{a}}\left( \frac{\pi}{2}-\arctan\left( \frac{b\theta\left(
c\right) }{2\sqrt{ac}}\right) \right) .  \label{a8}
\end{equation}
Returning then to the definitions in (\ref{a4}) and considering $G^{+}$ we
now define the variables $u$ and $v$ by changing the sign respectively: $%
u=y+x$ and $v=x.$This results in the expression for the $I^{+}\left(
q\right) $ which differs from $I^{-}\left( q\right) $ in (\ref{a6}) by the
change of sign in front of $b,$ so that integration over $v$ will differ
from (\ref{a8}) by the sign in front of the $\arctan$ function. These terms
then cancel each other in Eq. (\ref{a3}) when expressed in the $u$ and $v$
variables, so that our integral simplifies to%
\begin{equation}
I=\pi q\int_{-\infty}^{\infty}\left\vert
\int_{\sigma/2-1}^{\sigma/2}\exp\left( idk_{F}uz\right) zdz\right\vert ^{2}%
\frac{du}{\sqrt{Q^{2}+u^{2}}}.  \label{a9}
\end{equation}
The $z-$ integral can also be integrated exactly, so that after the
replacement of variable $x=udk_{F}$, Eq. (\ref{a9}) becomes%
\begin{equation}
I=\int_{-\infty}^{\infty}\frac{2\sigma\left( \sigma-2\right) x^{2}\left(
1-\cos x\right) +4x^{2}+8\left( 1-\cos x-x\sin x\right) }{4x^{4}\sqrt{\left(
dq\right) ^{2}+x^{2}}}dx.  \label{a10}
\end{equation}
The last integral can be reduced to the Meijer G-function \cite{Olver},
however for the long wavelength approximation $dq<1$ one can also obtain
directly from (\ref{a10}) its excellent approximation for $dq<1$ by the
following expression 
\begin{equation}
I\left( q\right) \simeq\pi q\left( -\frac{\left( \sigma-1\right) ^{2}}{2}%
\ln\left( dq\right) +0.81(\sigma-1)^{2}+\frac{1}{8}\right) .  \label{a11}
\end{equation}

\end{document}